\documentclass[iop]{emulateapj} 
\usepackage{amsmath}

\newcommand{\dwarf}{J1229+02}

\newcommand{\galex}{{\sl GALEX}}   
\newcommand{\Ha}{\ensuremath{{\rm H}\alpha}}   
\newcommand{\HI}{\ion{H}{1}}   
\newcommand{\hst}{{\sl HST}} 
\newcommand{\kms}{\ensuremath{{\rm km\,s}^{-1}}}
\newcommand{\lya}{\ensuremath{{\rm Ly}\alpha}}

\begin{document}

\title{Updated Models for the Creation of a Low-$z$ QSO Absorber by a
       Dwarf Galaxy Wind\altaffilmark{1}}

\author{Brian A. Keeney, Peter Joeris, John T. Stocke, Charles
        W. Danforth, and Emily M. Levesque\altaffilmark{2}}

\affil{Center for Astrophysics and Space Astronomy, Department of
       Astrophysical and Planetary Sciences, University of Colorado,
       \\   389 UCB, Boulder, CO 80309, USA; brian.keeney@colorado.edu}

\altaffiltext{1}{Based on observations with the NASA {\sl Galaxy
       Evolution Explorer} (\galex). \galex\ is operated for NASA by
       the California Institute of Technology under NASA contract
       NAS5-98034.}  \altaffiltext{2}{Hubble Fellow}

\shorttitle{Updated Models for Absorber Creation} \shortauthors{Keeney
et~al.}
\submitted{Accepted for publication in \textit{The Astronomical Journal}.}

\begin{abstract}
We present new \galex\ images and optical spectroscopy of \dwarf, a
dwarf post-starburst galaxy located 81~kpc from the 1585~\kms\
absorber in the 3C~273 sight line. The absence of \Ha\ emission and
the  faint \galex\ UV fluxes confirm that the galaxy's recent star
formation rate is $<10^{-3}~M_{\Sun}\,{\rm  yr^{-1}}$. Absorption-line
strengths and the UV-optical SED give similar estimates of the
acceptable model parameters for its youngest stellar population where
$f_m<60$\% of its total stars (by mass) formed in a burst $t_{\rm
sb}=0.7$--3.4~Gyr ago with a stellar metallicity of $-1.7 < {\rm
[Fe/H]} < +0.2$; we also estimate the stellar mass of \dwarf\ to be
$7.3 < \log{(M_*/M_{\Sun})} < 7.8$. Our previous study of \dwarf\
found that a supernova-driven wind was capable of expelling all of the
gas from the galaxy (none is observed today) and could by itself
plausibly create the nearby absorber. But, using new data, we find a
significantly higher galaxy/absorber velocity difference, a younger
starburst age, and a smaller starburst mass than previously
reported. Simple energy-conserving  wind models for \dwarf\ using
fiducial values of $f_m\sim0.1$, $t_{\rm sb}\sim2$~Gyr, and
$\log{(M_*/M_{\Sun})}\sim7.5$ allow us to conclude that the galaxy
alone cannot produce the observed QSO absorber; i.e., any putative
ejecta must interact with ambient gas from outside \dwarf.  Because
\dwarf\ is located in the southern extension of the Virgo cluster
ample potential sources of this ambient gas exist. Based on the two
nearest examples of  strong metal-line absorbers discovered
serendipitously (the current one and the 1700~\kms\ metal-line
absorber in the nearby Q1230+0115 sight line), we conclude that
absorbers with $10^{14}<N_{\rm H\,I}<10^{16}~{\rm cm^{-2}}$ at impact
parameters $\gtrsim1\,R_{\rm vir}$ are likely intergalactic systems
and cannot be identified unambiguously as the circumgalactic material
of any one individual galaxy.
\end{abstract}


\section{Introduction}
\label{intro}

The low-redshift ``\lya\ forest'' was discovered by \citet{morris91}
and \citet{bahcall91} in the ultraviolet spectrum of 3C~273 using the
first generation spectrographs on the {\sl Hubble Space Telescope}
(\hst). Two specific \lya\ + metal-line absorbers\footnote{The
1015~\kms\ absorber will not be discussed further in this Paper; see
\citet{stocke13,stocke14} and \citet{savage14} for new analysis of
this absorber.} at $cz=1015$ and 1585~\kms\ in the spectrum of 3C~273
attracted significant early and continuing interest \citep{morris91,
weymann95, sembach01, tripp02, rosenberg03} due to their high column
density ($\log{N_{\rm H\,I}} = 14.41$ and 15.85~cm$^{-2}$,
respectively) and proximity to Earth.

Extensive galaxy surveys in the region around 3C~273 \citep{morris91,
morris93} and other \hst-observed targets
\citep*[e.g.,][]{tripp98,prochaska04,prochaska11,chen09,johnson13}
have found that, in general, the association between absorbers and
galaxies is rather loose. Most low column density ($\log{N_{\rm H\,I}}
< 14.0~{\rm cm}^{-2}$) absorbers are associated not with individual
galaxies but rather with large-scale ``filaments'' of galaxies. A
small percentage ($\sim20$\%) of these low column density absorbers
are even found in galaxy voids \citep{stocke07}.

At higher \HI\ column densities, \lya\ absorbers are routinely found
associated with individual galaxies or groups of galaxies
\citep{stocke13, stocke14}. At $\log{N_{\rm  H\,I}} \geq 14.5~{\rm
cm}^{-2}$, approximately 50\% of \lya\ absorbers are projected within
the virial radius of a galaxy, almost always a star-forming galaxy
\citep{stocke13,tumlinson11,thom12}.  The percentage of absorption
systems closely associated with galaxies increases with column
density. While these associations are usually with $L \geq 0.1\,L^*$
galaxies, a smaller proportion are associated with dwarfs
\citep{stocke13}.  For Lyman-limit systems (LLS; $\log{N_{\rm H\,I}}
\geq 17.2~{\rm cm}^{-2}$) virtually all absorber systems are found
within $\approx100$~kpc of a $L \geq 0.1\,L^*$ galaxy
\citep{steidel93}. At the highest column densities ($\log{N_{\rm
H\,I}} \geq 20.3~{\rm cm}^{-2}$), the damped \lya\ absorbers (DLAs)
are associated with galaxies with a variety of luminosities
\citep*{wolfe05,zwaan05,rosenberg03a} and are thought to arise
primarily in the thick gaseous disks of galaxies \citep{wolfe05} but
some of these systems may arise in tidal debris between galaxies.

Given the previous results on absorber-galaxy associations and the
relatively high column density of the 1585~\kms\ absorber, we might
expect to find a nearby galaxy for which the 3C~273 sight line is
projected within its virial radius. Indeed, a post-starburst dwarf
spheroidal galaxy (SDSS~J122950.57+020153.7, \dwarf\ hereafter) was
discovered only 71~kpc away on the sky \citep{morris93}. Not only is
\dwarf\ the closest galaxy to the absorber but also the post-starburst
spectrum is unusual for such a small galaxy ($M_B =-16$).
\citet[Paper~1 hereafter]{stocke04} showed that a wind emanating from
this dwarf due to a recent starburst is plausibly sufficient to
produce the 1585~\kms\ \lya\ absorber. This putative wind also appears
to have left the system without any detectable gas
\citep[Paper~1;][]{vangorkom93}, which suggests that \dwarf\ will
simply fade over the next few billion years to eventually attain a
luminosity comparable to the most luminous Local Group dwarfs unless
it accretes new gas to fuel subsequent star formation episodes. This
scenario is reminiscent of galaxy evolution for the ``faint blue
galaxy'' population proposed by \citet{babul92}.

This hypothesis was proposed to explain the plethora of very faint,
blue galaxies found by deep CCD imaging surveys
\citep[$B\sim24$--27;][]{tyson88}. A star-bursting dwarf at $M_B \sim
-18$ would have $B\sim24$ at $z\sim0.5$ and very blue colors. But if a
supernova-driven wind cleared almost all of the gas from this small
galaxy, then it would form no more stars and simply fade. Eventually
such a galaxy could reach the very low luminosity of a Local Group
dwarf ($M_B \geq -14$) if there was no new star formation. Without a
significant fading of the bulk of the very numerous faint, blue galaxy
population at $z\gtrsim0.5$, it is not possible to understand their
absence in the current epoch. \dwarf\ is the best current example of a
\citeauthor{babul92} object in the process of fading.

If the galaxy wind interpretation for \dwarf\ is correct, this
absorber/galaxy pair may also be the best example of a dwarf galaxy
wind producing a metal-enriched absorber due to gas that is escaping
into the intergalactic medium (IGM). The absorber metallicity of only
6\% solar roughly matches the metallicity obtained for \dwarf\ using
Lick indices ($[{\rm Fe/H}] = -1.0\pm0.5$; see Paper~1). For a
single-burst model \citep{poggianti01, bruzual93, worthy94} in which
virtually all of the luminosity of the galaxy was created, an age of
$3.5\pm1.5$~Gyr is obtained for the starburst. If a significant number
of older stars is present in the galaxy, then a smaller, younger
starburst is required to obtain the observed absorption line
strengths, which roughly match the spectral signature of late-A and
early-F type main sequence stars. An overabundance of [Si/C] suggests
an absorber origin due to recent Type~II supernovae explosions
\citep[past few billion years;][]{tripp02}. The metallicity values and
silicon overabundance support the association of the absorber and
\dwarf.

Based on the single-burst model for \dwarf, the galaxy/absorber radial
velocity difference and the luminosity-derived galaxy mass, Paper~1
constructed a simple model confirming that the 1585~\kms\ absorber
could have been produced by a supernova-powered wind from  \dwarf,
even in the case where the mass outflow from the galaxy is entirely
isotropic. However, there were a few uncertainties involved in this
modeling, including whether the galaxy/absorber radial velocity
difference reported in Paper~1 was correct and whether a single-burst
model is viable given the very high star formation rate required
($\sim 1~M_{\Sun}\,{\rm yr}^{-1}$; Paper~1).

\begin{figure*}[!t]
\epsscale{1.00} \centering \plotone{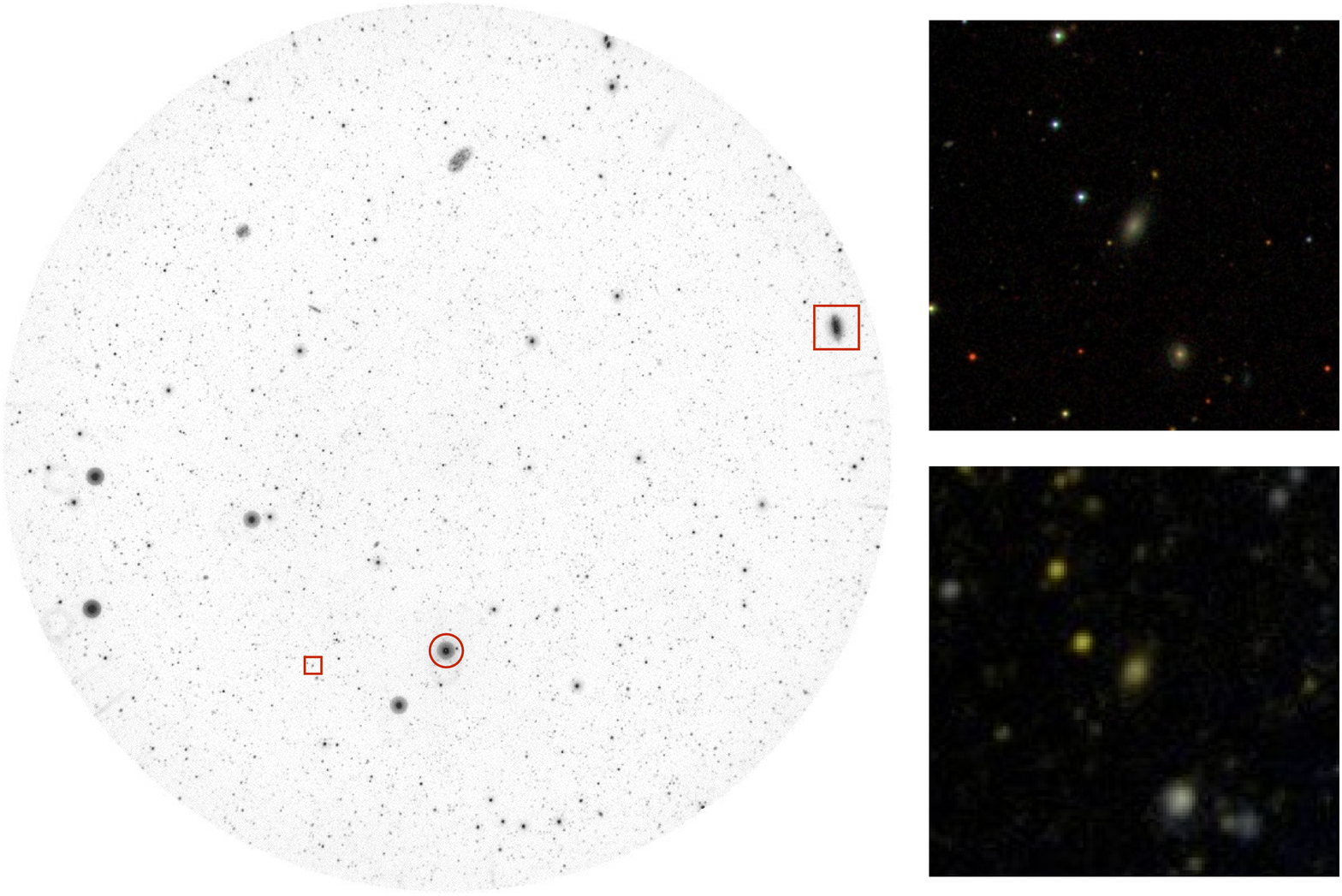}
\caption{SDSS and \galex\ images of \dwarf. At left we show \galex\
tile GI4\_012003 (1\fdg2 diameter), which we use for UV photometry in
Section~\ref{galex}. A circle indicates the position of 3C~273 and
boxes indicate the positions of NGC~4409 (big box) and \dwarf\ (small
box). At right we show zoomed-in color ($\sim3\arcmin\times3\arcmin$
FOV) images centered on \dwarf\ from SDSS (top) and \galex\ (bottom).
\label{fig:images}}
\end{figure*}

However, even if these modelling uncertainties can be resolved we are
left with the fundamental ambiguity of associating the 1585~\kms\
absorber with any individual galaxy. \dwarf\ is located in the
southern extension of the Virgo cluster where there are many galaxies
at similar velocities (see also Figure~2 of \citealp{stocke14} for the
distribution of galaxies near 3C~273 at Virgo velocities). Given the
faint completeness limits of the Sloan Digital Sky Survey (SDSS) and
pointed surveys in the region \citep[e.g.,][]{morris93}, there is
little doubt that \dwarf\ is the closest galaxy \citep[$\sim1\,R_{\rm
vir}$ in projection;][]{stocke13} to the 3C~273 sight line with a
velocity near that of the 1585~\kms\ absorber. However, due to its
higher luminosity NGC~4409 is a comparable number of virial radii from
the 3C~273 sight line \citep[$\sim1.5\,R_{\rm vir}$ in projection,
although low-ionization absorbers like the one we study are typically
associated with galaxies located within $\sim0.6\,R_{\rm
vir}$;][]{stocke13} despite having a projected physical distance
$\sim3.5$ times larger than that of \dwarf\ \citep{stocke13}. The
recession velocity of NGC~4409 is also closer to the absorber velocity
than that of \dwarf. In Figure~\ref{fig:images} we show the locations
of 3C~273, NGC~4409, and \dwarf\ in the \galex\ tile GI4\_012003, as
well as zoomed-in SDSS and \galex\ views of \dwarf. Thus, the density
of galaxies in this part of the sky is such that we cannot
unambiguously associate the 1585~\kms\ absorber in the 3C~273 sight
line with any individual galaxy, and an association with larger-scale
group/cluster gas is also a distinct possibility
\citep{yoon12,stocke14}.

Nevertheless, the star formation history of \dwarf\ and its connection
to the \citet{babul92} picture of fading blue galaxies make it an
interesting object to study in its own right. In this Paper we present
new high signal-to-noise optical spectroscopy (Section~\ref{apo}) and
new {\sl Galaxy Evolution Explorer} (\galex) images
(Section~\ref{galex}) of \dwarf. The revised recessional velocity of
\dwarf\ increases the velocity difference between the galaxy and the
absorber, increasing the energy required to eject the absorber to its
observed distance from the dwarf if the absorber originates in a
galaxy wind. Despite the increased energy requirements, we show in
Section~\ref{model} that a simple model still fits this
absorber/galaxy system and argue that this tiny dwarf could plausibly
create the strong nearby absorber. Our results are summarized in
Section~\ref{conclusion}.

\section{Optical Spectroscopy of \dwarf}
\label{apo}

New high-resolution optical spectra of \dwarf\ were obtained with the
Dual Imaging Spectrograph (DIS) at the Apache Point Observatory (APO)
3.5-m telescope to better constrain the lack of \Ha\ emission and
refine the galaxy's recession velocity, which SDSS ($1775\pm12$~\kms)
found to be discrepant from the value in Paper~1
($1635\pm50$~\kms). Contamination of the galaxy's \ion{Ca}{2}~K by
solar \ion{Ca}{2}~H absorption was suspected in the Las~Campanas 2.6-m
spectrum used in Paper~1 due to the presence of moonlight during the
observation; therefore, we were careful to observe \dwarf\ only during
dark time at APO.

\dwarf\ was observed for 3~hours on 2011~May~05 with the R1200 grating
and 2~hours on 2011~May~25 with the B1200 and R1200 gratings. A
$1\farcs5$-wide slit was used on both nights and both gratings have
dispersions of $\approx0.6~{\rm arcsec~pix^{-1}}$, corresponding to
velocity resolutions of $\sim70$~\kms\ and $\sim100$~\kms\ at the
observed positions of \Ha\ and H$\beta$, respectively. We observed
\dwarf\ with the highest resolution gratings for which it was feasible
to maximize our ability to resolve \Ha\ emission, but the resulting
spectra only cover wavelength regions of 5800--6900~\AA\ with ${\rm
S/N} \approx 10~{\rm pix^{-1}}$ and 4150--5400~\AA\ with ${\rm S/N}
\approx 5~{\rm pix^{-1}}$. Thus, the SDSS spectrum is superior to ours
in terms of both wavelength coverage and S/N and we defer to it in our
analysis in situations where these properties are advantageous.

\begin{figure}[!t]
\epsscale{1.00} \centering \plotone{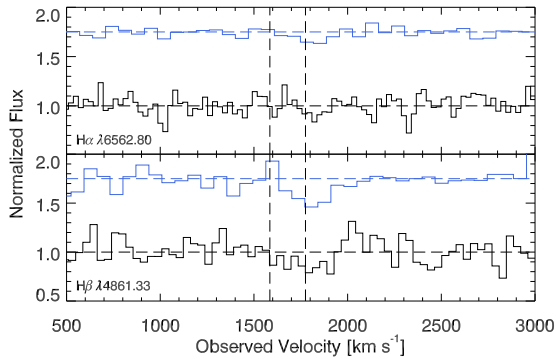}
\caption{High-resolution optical spectrum of \dwarf\ from APO/DIS
  (black) compared to the SDSS spectrum (blue) in the regions of \Ha\
  and H$\beta$ absorption from \dwarf. The QSO absorber velocity of
  1585~\kms\ and the SDSS-derived galaxy velocity of 1775~\kms\ are
  indicated by the dashed vertical lines. In the APO/DIS spectrum, one
  pixel corresponds to $\sim25$~\kms\ at \Ha, and $\sim40$~\kms\ at
  H$\beta$.
\label{fig:DIS}}
\end{figure}

Figure~\ref{fig:DIS} shows the APO/DIS spectrum compared to the lower
resolution ($\sim150$~\kms) SDSS spectrum in the regions of \Ha\ and
H$\beta$ absorption from the dwarf galaxy. The two spectra are
qualititatively very similar, and the higher resolution of the APO/DIS
spectrum reveals no evidence of \Ha\ emission. There is also no
evidence of a difference in recession velocity between the two spectra
on the basis of weak \Ha\ and H$\beta$ absorption, the only lines
(marginally) detected in our APO/DIS spectrum, so we adopt the SDSS
value of $1775\pm12$~\kms\ as final. This means that the line of sight
velocity difference between the galaxy and absorber has increased to
$190\pm13$~\kms\ from the $50\pm50$~\kms\ value reported in
Paper~1. The absorber velocity of $1585\pm3$~\kms\ and the galaxy
velocity are indicated by the dashed vertical lines in
Figure~\ref{fig:DIS}.

Estimates of the galaxy's age and metallicity in Paper~1 were
determined from an analysis of its Lick indices, particularly the
${\rm H_{\gamma}^F}$ and H$\beta$ indices for age sensitivity and the
${\rm \langle Fe \rangle}$ and ${\rm Mg_2}$ indices for metallicity
sensitivity \citep{poggianti01}. In the meantime, considerable effort
has gone into recalibrating the Lick indices at SDSS resolution
\citep[e.g.,][]{kauffmann03,eisenstein03,bruzual03,schiavon07,graves08,
franchini10,franchini11} so their power can be brought to bear on the
vast array of SDSS spectra. We have used the EZ\_Ages software of
\citet{graves08} to measure Lick indices from the SDSS spectrum of
\dwarf\ and calculate its age and metallicity. This analysis finds an
age of $2.1^{+1.0}_{-0.5}$~Gyr and a metallicity of $[{\rm Fe/H}] =
-1.2^{+\infty}_{-0.5}$, which despite the lack of a formal upper bound
on metallicity is nonetheless broadly consistent with the age
($3.5\pm1.5$~Gyr) and metallicity ($[{\rm Fe/H}] = -1.0\pm0.5$) of
Paper~1.

\section{\galex\ Imaging of \dwarf}
\label{galex}

\begin{deluxetable*}{lccccccc}

\tablecolumns{8}
\tablewidth{0pt}

\tablecaption{\galex\ and SDSS Photometry of \dwarf
\label{tab:photometry}}

\tablehead{ \colhead{Filter} & \colhead{$\lambda_{\rm p}$\tablenotemark{a}} & \colhead{$\Delta\lambda_{\rm rect}$\tablenotemark{a}} & \colhead{Pipeline Mag.\tablenotemark{b}} & \colhead{$A_{\lambda}$ (mag)} & \colhead{Corrected Mag.\tablenotemark{c}} & \colhead{$\log{F}$\tablenotemark{d}} & \colhead{$\log{L}$\tablenotemark{e}}}

\startdata
FUV &  1535 &  255 & $22.229\pm0.034$ & $0.135\pm0.012$ & $22.16\pm0.06$ & $-13.79\pm0.02$ & $39.10\pm0.05$ \\
NUV &  2301 &  730 & $21.110\pm0.010$ & $0.147\pm0.009$ & $20.96\pm0.03$ & $-13.21\pm0.01$ & $39.69\pm0.02$ \\
$u$ &  3557 &  558 & $18.674\pm0.050$ & $0.082\pm0.001$ & $18.59\pm0.10$ & $-12.76\pm0.04$ & $40.14\pm0.07$ \\ 
$g$ &  4702 & 1158 & $17.284\pm0.007$ & $0.064\pm0.001$ & $17.22\pm0.03$ & $-12.13\pm0.01$ & $40.76\pm0.03$ \\
$r$ &  6176 & 1111 & $16.733\pm0.007$ & $0.044\pm0.001$ & $16.69\pm0.04$ & $-12.17\pm0.02$ & $40.72\pm0.03$ \\ 
$i$ &  7490 & 1045 & $16.481\pm0.008$ & $0.033\pm0.001$ & $16.45\pm0.03$ & $-12.27\pm0.01$ & $40.62\pm0.03$ \\ 
$z$ &  8947 & 1125 & $16.286\pm0.026$ & $0.024\pm0.001$ & $16.26\pm0.06$ & $-12.32\pm0.02$ & $40.57\pm0.04$ 
\enddata

\tablenotetext{a}{Filter pivot wavelength and rectangular width in units of \AA, calculated using the filter's effective area \citep[\galex;][]{morrissey07} or photometric response \citep*[SDSS;][]{doi10} curves.}
\tablenotetext{b}{Pipeline FUV and NUV magnitudes from \galex\ and optical model magnitudes from SDSS.}
\tablenotetext{c}{Pipeline magnitudes corrected for extinction and photometric calibration uncertainties; see text for details.}
\tablenotetext{d}{Logarithm of integrated galaxy flux in the filter bandpass in units of ${\rm erg\,s^{-1}\,cm^{-2}}$.}
\tablenotetext{e}{Logarithm of integrated galaxy luminosity in the filter bandpass in units of ${\rm erg\,s^{-1}}$.}

\end{deluxetable*}

The absorption-line analysis presented in the preceding Section and in
Paper~1 is one method of attempting to constrain the age and
metallicity of a galaxy. Another is to examine the galaxy's broadband
spectral energy distribution (SED) and model it as the superposition
of one or more simple stellar populations with single-valued ages and
metallicities. This endeavour is made easier if the galaxy to be
modelled is observed in its rest-frame ultraviolet where there can be
a large difference between the luminosity of a population of young and
old stars.

To this end, \dwarf\ was observed by \galex\ for 1.66~ksec in GI
Cycle~4 as part of tile GI4\_033001 (PI: B.~Keeney). The goal of these
observations was to measure the NUV and FUV magnitudes of \dwarf\ and
use them to model its star formation history.  Fortuitously,
additional \galex\ images of the galaxy were acquired in the same
Cycle as part of tile GI4\_012003 (PI: K.~Sembach). This tile is much
deeper than the one obtained as part of our program (FUV exposure time
of $\sim30$~ksec and NUV exposure time of $\sim50$~ksec), so we use it
for all subsequent analysis in this Paper.

\galex\ images in the FUV and NUV channels simultaneously with spatial
resolutions of $4\farcs2$ and $5\farcs3$, respectively, and a circular
field of view $1\fdg2$ in diameter \citep{morrissey07}. A
sophisticated data reduction pipeline is employed to reconstruct
images from the raw telemetry; many details of this pipeline can be
found in \citet{morrissey07}, but we briefly list its salient features
here. Photon positions are corrected for various instrumental effects
and turned into raw-count images, then divided by a relative response
image that incorporates the effective exposure time of each pixel to
create a flux-calibrated intensity image. The background in the
intensity image is estimated using a custom threshold image and then
subtracted. Finally, SExtractor \citep{bertin96} is used to determine
source positions and photometry in the NUV and FUV images and the NUV
and FUV source catalogs are combined to form a merged catalog of
source positions, fluxes, and magnitudes.

As evidenced by the zoomed-in image in the bottom right corner of
Figure~\ref{fig:images}, \dwarf\ has several faint neighbors within
$\sim1\arcmin$ in the deep \galex\ exposures so care is needed when
determining its magnitude.  Despite its low redshift, \dwarf\ is small
enough on the sky to be only marginally resolved by \galex\ (90\% of
its $r$-band flux is contained within a $10\arcsec$ radius; see
Figure~\ref{fig:images}). Thus, we effectively treat \dwarf\ as a
point source in the \galex\ images and use aperture magnitudes
(specifically, the ``APER\_4'' magnitudes with a radius of $8~{\rm
pix} = 12\arcsec$) chosen to have a size that roughly corresponds to
the size of the galaxy in the SDSS images.  This procedure yields an
NUV magnitude for \dwarf\ of $21.11\pm0.01$ and an FUV magnitude of
$22.30\pm0.03$.  For comparison, the NUV and FUV magnitudes of \dwarf\
derived from our notably shallower tile are $21.08\pm0.06$ and
$22.43\pm0.17$, respectively.  We also note that the UV surface
brightness profile of \dwarf\ is quite smooth, consistent with its
optical surface brightness profile and lack of current star formation
(Paper~1).

\subsection{UV-Optical SED of \dwarf}
\label{galex:SED}

\begin{deluxetable*}{lccccccccccc}

\tablecolumns{12}
\tablewidth{0pt}

\tablecaption{Parameters from SED Fits to \dwarf
\label{tab:SED}}

\tablehead{                          && \multicolumn{4}{c}{\textit{Single-burst Models}}                                                         && \multicolumn{5}{c}{\textit{Two-population Models}} \\
            \cline{3-6} \cline{8-12} \\[-1ex]
            \colhead{${\rm [Fe/H]}$} && \colhead{$\chi_{\nu}^2$} & \colhead{$\mathcal{L}$\%} & \colhead{Age} & \colhead{$\log{M_*}$} && \colhead{$\chi_{\nu}^2$} & \colhead{$\mathcal{L}$\%} & \colhead{$f_m$} & \colhead{Age} & \colhead{$\log{M_*}$} }

\startdata
$-1.54$ &&  6.0 &    62.5 & $3.10^{+0.19}_{-0.25}$ & $7.48^{+0.02}_{-0.03}$ && 4.8 &    59.2 & $0.15^{+0.44}_{-0.11}$ & $2.64^{+0.47}_{-0.63}$ & $7.58^{+0.17}_{-0.09}$ \\[1ex]
$-0.54$ &&  5.8 &    37.3 & $1.27^{+0.04}_{-0.02}$ & $7.27^{+0.00}_{-0.01}$ && 4.3 &    20.7 & $0.05^{+0.42}_{-0.04}$ & $1.04^{+0.24}_{-0.15}$ & $7.49^{+0.29}_{-0.22}$ \\[1ex]
$-0.24$ &&  7.2 & \phn0.3 & $0.91^{+0.05}_{-0.01}$ & $7.24^{+0.01}_{-0.01}$ && 4.1 &    16.7 & $0.03^{+0.06}_{-0.02}$ & $0.77^{+0.09}_{-0.09}$ & $7.61^{+0.15}_{-0.13}$ \\[1ex]
$+0.15$ &&  9.1 &  $<0.1$ & $0.62^{+0.02}_{-0.02}$ & $7.17^{+0.01}_{-0.01}$ && 4.4 & \phn2.5 & $0.01^{+0.02}_{-0.00}$ & $0.46^{+0.08}_{-0.02}$ & $7.67^{+0.07}_{-0.08}$ \\[1ex]
$+0.55$ &&  9.5 &  $<0.1$ & $0.49^{+0.02}_{-0.01}$ & $7.12^{+0.01}_{-0.01}$ && 4.5 & \phn0.9 & $0.01^{+0.01}_{-0.00}$ & $0.38^{+0.04}_{-0.02}$ & $7.70^{+0.02}_{-0.06}$ 
\enddata

\end{deluxetable*}

The UV-optical SED of \dwarf\ was constructed from its \galex\
aperture magnitudes and its SDSS model magnitudes.
Table~\ref{tab:photometry} lists the wavelength and width of each
filter, along with the pipeline magnitudes, total extinction,
extinction-corrected magnitude, and integrated flux and luminosity for
\dwarf\ in each of the \galex\ and SDSS bands.

Galactic foreground extinction was determined using the extinction law
of \citet{fitzpatrick99} assuming $R_V = 3.1$ and $E(B-V) =
0.017\pm0.010$~mag \citep{schafly11}. \dwarf\ has no optical emission
lines (Paper~1; Section~\ref{apo}) and no \HI\ 21-cm emission
\citep[$M_{\rm H\,I} < 5\times10^6~M_{\Sun}$;][]{vangorkom93}, so
there is no evidence of gas or dust in this object; thus, we assume
that it has no intrinsic extinction. 3C~273 has a large Galactic
latitude ($b\approx75\degr$) and a correspondingly low color excess so
that the large fractional errors associated with extrapolating
extinction in the \galex\ bands still yield modest absolute
uncertainties.

In the sixth column of Table~\ref{tab:photometry} we list the
extinction-corrected magnitude in each filter. The uncertainty in this
quantity is the quadrature sum of the error in the extinction
correction and the absolute photometric uncertainty interpolated from
the results of \citet[where it is called
``repeatability'']{morrissey07} for \galex\ and \citet{strateva01} for
SDSS. These larger uncertainties are propagated through to the
uncertainty in the integrated flux, but when calculating the
integrated luminosity we use the $2\sigma$ uncertainty in each
filter. We do this as an ad hoc way of accounting for: (1) the fact
that the models to which we are comparing the data are not perfect
(i.e., they have errors of their own that are hard to quantify); (2)
there may be systematic offsets in the zero-point calibration between
the \galex\ and SDSS data; and (3) we may have systematically
underestimated the \galex\ magnitudes relative to SDSS by choosing to
use aperture magnitudes.

When calculating the luminosity in each band we assume that \dwarf\ is
located at a distance of 25.6~Mpc, the luminosity distance at
$cz=1775$~\kms; however, the distances to galaxies in this part of the
Virgo cluster are known to be double-valued \citep{tonry81}. For
example, the distance to the nearby giant elliptical galaxy NGC~4636
has been determined from surface brightness fluctuations and globular
cluster luminosity function values to be 16.4~Mpc (mean value given in
NED\footnote{The NASA/IPAC Extragalactic Database (NED) is operated by
the Jet Propulsion Laboratory, California Institute of Technology,
under contract with the National Aeronautics and Space
Administration.}); clearly if \dwarf\ is at a similar distance it will
have a smaller luminosity (by $\approx0.4$~dex). However, since this
is a systematic effect it does not change any of our modeled
quantities except the galaxy's stellar mass, which needs to be revised
downward by the same factor of $\approx0.4$~dex.

In the next two subsections we present a detailed description of the
SED modeling (first single-burst and then two-population models) that
are allowed by the data. A synthesis of this modeling is presented in
Section~\ref{galex:summary}.

\subsubsection{Single-burst Models}
\label{galex:SED:1pop}

In order to model the SED of \dwarf\ we utilized a grid of template
SEDs from  Starburst99 \citep{leitherer99}. We used the original
Padova models at all available metallicities \citep[metal fractions of
0.0004, 0.004, 0.008, 0.02, and 0.05, corresponding to metallicities
of ${\rm [Fe/H]} = -1.54$, $-0.54$, $-0.24$, +0.15, and +0.55,
respectively;][]{asplund09} and assumed a  Salpeter initial mass
function (IMF) with masses ranging from 0.1--$100~M_{\Sun}$ in all
cases. The Padova models were chosen because they are the only models
available that have metal fractions below 0.001 \citep[i.e., ${\rm
[Fe/H]} < 1.1$;][]{asplund09}. For each metallicity we generated
templates with ages ranging from 0.1--10~Gyr in 10~Myr time steps. The
templates are generated for a simple stellar population with an
assumed mass of $10^6~M_{\Sun}$ and must be scaled to match the
observed luminosity of \dwarf; this scaling gives an estimate of the
galaxy's stellar mass.

To compare the template SEDs with the observed galaxy luminosities in
Table~\ref{tab:photometry}, we determined synthetic luminosities in
each of the filters by convolving the template SED at each grid point
with the filter response functions \citep{morrissey07,doi10}. Fitting
then proceeded in two steps. First, for each age at a given
metallicity we scaled the template luminosities assuming a grid of
stellar mass values ranging from $\log{(M_*/M_{\Sun})} = 6$--9 in
steps of 0.01~dex (hereafter any masses appearing in logarithms are
assumed to have units of solar masses) and determined $\chi_{\rm
opt}^2$, a $\chi^2$ value using only the optical ($g$, $r$, $i$, $z$)
data. Then we calculated $\chi_{\rm tot}^2$ for each age and
metallicity, the $\chi^2$ value between the observed and synthetic
luminosities calculated using all of the filters with the stellar mass
fixed at the value that minimized $\chi_{\rm opt}^2$ at that grid
point.  The bluest data points (FUV, NUV, and $u$-band) are crucial
for distinguishing models since the largest variations between younger
and older stellar populations occur at $\lambda \lesssim 4000$~\AA.

The parameters of \dwarf\ that we are hoping to estimate with this
model are the age, metallicity, and mass of its stellar population
under the assumption that all of the stars in the galaxy were formed
in a single burst. However, given that we pre-select stellar mass
values in our fitting procedure the resulting best-fit values of age
and metallicity are conditionally dependent on the choice of stellar
mass. The likelihood of our model parameters given the data is
$\mathcal{L}({\rm [Fe/H]}, t | \log{M_*}) \propto e^{-\chi_{\rm
tot}^2/2}$, from which we determine a highest posterior density
credible interval for the age of the population at a given
metallicity. We do this by finding the likelihood threshold above
which 95\% of the total likelihood lies and determining the range of
ages with likelihoods above that threshold. We interpret this result
as an indicative range of ages that produce plausible models.

Table~\ref{tab:SED} lists the results of our single-burst fits. For
each metallicity we list: (1) the reduced $\chi^2$ of the best-fitting
model ($\chi_{\nu}^2 \equiv \chi_{\rm tot}^2/7$); (2) the percentage
of the total likelihood found in models of that metallicity; (3) the
mode of the age distribution and indicative range of ages as described
above; and (4) the range of plausible stellar masses. To determine the
range of plausible stellar masses we found the $\log{M_*}$ values
chosen (i.e., values that minimize $\chi_{\rm opt}^2$ for a given age
and metallicity) by grid points that are above the 95\% likelihood
threshold. The values quoted in Table~\ref{tab:SED} indicate the
median, minimum, and maximum of this range of stellar masses.

There is clearly some degeneracy in the values in Table~\ref{tab:SED}
since higher metallicity models always prefer younger, less massive
bursts than lower metallicity models. Nevertheless, since $>99$\% of
the total likelihood in Table~\ref{tab:SED} is found in the two lowest
metallicity models, we infer that the single-burst SED fits prefer
models with ${\rm [Fe/H]} < -0.2$. Thus ages in the range
$\sim1.3$--3.4~Gyr and stellar masses of $\log{M_*} \sim 7.3$--7.5 are
plausible. It's unfortunate that we were unable to run
lower-metallicity grid points to ascertain whether the SED likelihood
continues to increase as the metallicity decreases; however, the
spectroscopic analysis of Paper~1 and Section~\ref{apo} are another
way of trying to model the galaxy as a single-burst population. The
ages inferred from these two techniques are very similar (we found an
age range of $2.1^{+1.0}_{-0.5}$~Gyr in Section~\ref{apo}) and the
Lick-index fits provided a lower-bound to the galaxy metallicity
(${\rm [Fe/H]} > -1.7$), while our SED fits provide a plausible upper
bound.

\begin{figure}[!t]
\epsscale{1.00} \centering \plotone{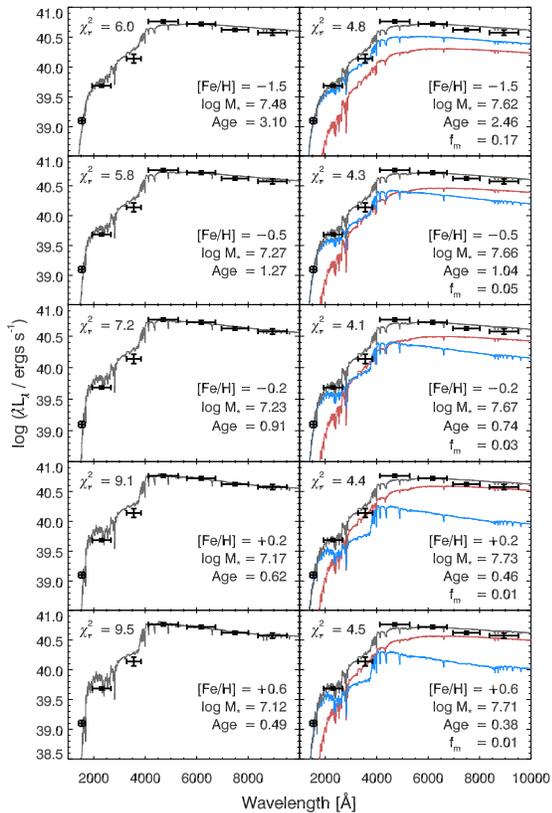}
\caption{Integrated luminosities of \dwarf\ in the \galex\ and SDSS
  bands (see Table~\ref{tab:photometry}) compared to template spectra
  from Starburst99. The left panels show the best-fitting single-burst
  models and the right panels show the best-fitting two-population
  models. Gray spectra are the templates that are compared to our
  observations. For the two-population models we also show the
  individual contributions of the younger (blue) and older (red)
  populations.
\label{fig:SED}}
\end{figure}

The left-hand panels of Figure~\ref{fig:SED} show the best-fitting
single-burst model for each metallicity. The reduced $\chi^2$ value
and stellar mass and age of the corresponding model grid point are
labelled in each panel. These parameter values, at which the joint
likelihood is maximized, are in good agreement with the maximal
marginal (i.e., individual) values listed in
Table~\ref{tab:SED}. Finally, we note that while the minimum
$\chi_{\nu}^2$ value in the entire grid occurs at a metallicity of
${\rm [Fe/H]} = -0.5$ a larger fraction of the points in the ${\rm
[Fe/H]} = -1.5$ grid have relatively small values of $\chi_{\nu}^2$;
hence, the lowest metallicity has 63\% of the total likelihood as
opposed to the 37\% that is associated with the metallicity of the
individual grid point that minimizes $\chi_{\nu}^2$
(Table~\ref{tab:SED}).

Both \Ha\ and FUV emission can be used to infer a galaxy's star
formation rate (SFR). Our APO/DIS spectrum of \dwarf\
(Section~\ref{apo}) yields $L(\Ha) < 2.4\times10^{36}~{\rm
erg\,s^{-1}}$ to $3\sigma$ significance and the \galex\ FUV luminosity
in Table~\ref{tab:photometry} corresponds to $\langle L_{\nu} \rangle
= (3.9\pm0.2)\times10^{24}~{\rm erg\,s^{-1}\,Hz^{-1}}$.  Using the
conversions of \citet*{hunter10} the \Ha\ SFR is
$<1.7\times10^{-5}~M_{\Sun}\,{\rm yr}^{-1}$ and the FUV SFR is
$(5.0\pm0.3)\times10^{-4}~M_{\Sun}\,{\rm yr}^{-1}$.  The higher SFR
inferred from the FUV flux is expected since \Ha\ measures star
formation over the past 10~Myr, whereas FUV measures star formation in
the past 10--100~Myr \citep{hunter10}. Furthermore, the FUV image
measures star formation over the entire galaxy, while the APO/DIS
spectrum only measures \Ha\ flux along the slit. However, even the
FUV-derived SFR of \dwarf\ is not appreciable, which is consistent
with the absorption-line and SED analyses summarized above.  If we
assume that the youngest stars in \dwarf\ formed in a burst that
lasted $\sim100$~Myr \citep{bruzual93} and created $\log{M_*} = 7.5$
solar masses of stars (i.e., as many stars as we could plausibly
explain in a single-burst model) the average SFR during the burst was
$\sim0.32~M_{\Sun}\,{\rm yr}^{-1}$, or $\sim600$ times the galaxy's
current SFR. This estimate of the galaxy's peak SFR is $\sim3$ times
less than the value derived in Paper~1 and is more consistent with
other dwarf starbursts.

\subsubsection{Two-population Models}
\label{galex:SED:2pop}

The single-burst models above provide acceptable fits to the galaxy
SED, but given the complex star formation histories inferred for Local
Group dwarf spheroidal galaxies \citep[e.g.,][]{mateo98,skillman05} it
is unlikely that all of the stars in \dwarf\ were formed in a single
burst.  However, we are limited in the complexity of star formation
models whose SEDs we can model by the small number of photometric data
points at our disposal.

To investigate the percentage of older stars that could be present in
\dwarf, we have created two-population SEDs combining an old (10~Gyr)
stellar population with ${\rm [Fe/H]} = -1.5$ and a younger population
with arbitrary age and metallicity. Under these assumptions we
introduce only one additional free parameter to our model: $f_m$, the
fraction of the galaxy's total mass that is contained in the younger
population. A related quantity necessary to determine the combined
luminosity of the two populations is $f_L$, the fraction of the
galaxy's light that is contained in the younger population:
\begin{equation}
f_L^{-1} = 1 + \frac{(M/L)_{\rm young}}{(M/L)_{\rm
old}}\left(\frac{1-f_m}{f_m}\right),
\label{eqn:fL}
\end{equation}
where $(M/L)_{\rm young}$ is the mass-to-light ratio of the younger
population and $(M/L)_{\rm old}$ is the mass-to-light ratio of the
older population. These values are clearly wavelength dependent so we
approximate them with average quantities and take advantage of the
fact that the Starburst99 template spectra are all generated for
populations with a fixed mass of $10^6~M_{\Sun}$. Thus, we simplify
Equation~\ref{eqn:fL} using
\begin{equation}
\frac{(M/L)_{\rm young}}{(M/L)_{\rm old}} \approx \frac{\langle\lambda
L_{\lambda}\rangle_{\rm old}}{\langle\lambda L_{\lambda}\rangle_{\rm
young}},
\label{eqn:ML}
\end{equation}
where $\langle\lambda L_{\lambda}\rangle_{\rm old}$ and
$\langle\lambda L_{\lambda}\rangle_{\rm young}$ are the average model
luminosities for the older and younger populations, respectively, in
the wavelength range 1000--10000~\AA.

Fitting was performed in much the same way as for the single-burst
model except that the grid now includes $f_m$ (ranging from 0.01--0.99
in steps of 0.01) in addition to age, metallicity, and stellar
mass. As in the single-burst case we first optimize stellar mass using
only the optical data before determining $\chi_{\rm tot}^2$ from all
of our data using the optimal stellar mass for a given grid point from
the first fitting step. The additional dimension of the two-population
model requires us to marginalize the likelihood over all possible
values of $f_m$ to determine a credible interval on age, and vice
versa. Plausible values of the stellar mass are determined in the same
way as for the single-burst model.

The results of our fits are listed in Table~\ref{tab:SED}. The
majority of the total likelihood is associated with the lowest
metallicity model, as in the single-burst case; similarly, the grid
point with the smallest $\chi_{\nu}^2$ has a metallicity (${\rm
[Fe/H]} = -0.2$) that is different from the metallicity with the
highest percentage of the total likelihood. There are some differences
between results of the two classes of models, however.

The reduced $\chi^2$ values for the two-population models are both
lower and more uniform than the corresponding values for the
single-burst models, and there is significantly more likelihood in the
${\rm [Fe/H]} = -0.2$ grid points for the two-population models than
for the single-burst models.  The plausible age distributions for the
two-population models decrease appreciably as compared to single-burst
models of the same metallicity and the stellar masses of the
two-population models are systematically higher than those of the
single-burst models.

Two-population models with higher metallicity prefer smaller mass
fractions for the younger population, smaller ages for this
population, and (generally) larger total stellar masses than lower
metallicity models.  More than 96\% of the total likelihood is
associated with the three lowest metallicity models, implying that the
two-population SED fits prefer metallicities of ${\rm [Fe/H]} <
+0.2$. The models at these metallicities have plausible parameter
values of $\log{M_*}\sim7.3$--7.8, $f_m \lesssim 0.6$, and age for the
younger population of $\sim0.7$--3.1~Gyr (Table~\ref{tab:SED}).

The right-hand panels of Figure~\ref{fig:SED} show the best-fitting
two-population model for each metallicity, with the reduced $\chi^2$,
stellar mass, age, and mass fraction of the corresponding grid point
labelled in each panel. Unlike the single-burst model, the points that
maximize the joint likelihood are not always in good agreement with
the modes of the marginal distributions; in particular, the stellar
mass that minimizes the reduced $\chi^2$ is remarkably constant and
usually at the high end of the plausible range listed in
Table~\ref{tab:SED}.

The results presented in Table~\ref{tab:SED} and Figure~\ref{fig:SED}
for two-population models assume a fixed age and metallicity of the
older stellar population in \dwarf\ (10~Gyr and ${\rm [Fe/H]} = -1.5$,
respectively). To test the robustness of the younger population
parameters to this assumption,  we have also run models with different
assumed values for the fixed age and metallicity of the older
population. If the older population is assumed to have the same
metallicity as in the models presented above but an age of 7.5~Gyr
instead of 10~Gyr, we find nearly identical values for $f_m$ and the
stellar mass of \dwarf\ as those presented in Table~\ref{tab:SED}, and
the inferred age of the younger population decreases by $\sim5$\%. If
the age of the older population is fixed at 5~Gyr we begin to find
more differences: the total stellar mass of the galaxy is
$\sim0.1$~dex smaller and the inferred ages are $\sim20$\% younger
than the values in Table~\ref{tab:SED}, and the mass fraction peaks at
$<5$\% for all metallicities. As the age of the older population
decreases the fraction of the total likelihood associated with a
metallicity of ${\rm [Fe/H]} = -1.5$ increases to $\sim80$\%, and the
fraction at metallicities of ${\rm [Fe/H]} = -0.5$ and $-0.2$ decrease
to $\sim10$\% and $\sim5$\%, respectively. Nevertheless, $>98$\% of
the total likelihood is associated with these three metallicities
regardless of the assumed age of the older population.

We also examined two-population models where the metallicity of the
older population was fixed at ${\rm [Fe/H]} = -0.5$ instead of ${\rm
[Fe/H]} = -1.5$, and had fixed ages of 10, 7.5, and 5~Gyr as above. We
did not intend this particular suite of models to be physically
consistent but rather to test the robustness of the values in
Table~\ref{tab:SED} and Figure~\ref{fig:SED} to the assumed
metallicity of the older population, so we allowed the younger
population to have any metallicity in our grid (i.e., we allowed the
metallicity of the younger population to be smaller than that of the
older population). We still find that $>90$\% of the total likelihood
is associated with younger populations having the three lowest
metallicities in our grid, but the fractions at each of these
metallicity values are nearly equal (i.e., $\sim30$\% of the total
likelihood resides in the unphysical model) so no individual
metallicity value is clearly preferred. When we compare these models
to the models described above where the older population has the same
fixed age but a lower metallicity, we find that the higher metallicity
models show very little difference in the total galaxy stellar mass
but prefer grid points with a higher mass fraction in the young
population ($f_m \sim 30$--50\% when the younger population has ${\rm
[Fe/H]} = -1.5$, and $f_m \sim 5$--20\% for all other metallicities)
and a larger age of the young population ($\sim10$--30\% larger at all
metallicities). It seems as if we are seeing the same degeneracy
manifest itself in the older population properties as we do in the
younger population, where assuming a younger, more metal-enriched
older population will lead to similar inferences about the properties
of the younger population as assuming an older, less metal-enriched
older population.

\subsection{Summary of SED Fitting}
\label{galex:summary}

In the preceding subsections we presented the methodology and results
of two classes of SED fits performed to \dwarf. The single-burst
models (Section~\ref{galex:SED:1pop}) assumed that all of the galaxy's
light was produced by a single stellar population whose age, mass, and
metallicity we attempted to estimate. The two-population models
(Section~\ref{galex:SED:2pop}) assumed an underlying older population
of fixed age and metallicity and attempted to estimate the age, mass
fraction, and metallicity of the younger population and the combined
mass of the two populations. We found that the simpler single-burst
models preferred metallicities of ${\rm [Fe/H]} < -0.2$, ages of
$\sim1.3$--3.4~Gyr, and stellar masses of $\log{M_*}\sim7.3$--7.5 for
\dwarf\ (Table~\ref{tab:SED}), consistent with the values found from
the absorption-line analyses of Paper~1 and Section~\ref{apo}.

Adding a second population improves the reduced $\chi^2$ of the
best-fit models, but interpreting the two-population models is more
complicated since any properties of the younger population depend in a
degenerate manner on the assumed properties of the older
population. Nevertheless, if we choose an older population with an age
of 10~Gyr and a metallicity of ${\rm [Fe/H]} = -1.5$ we find that the
younger population has a mass fraction of $<60$\%, a metallicity of
${\rm [Fe/H]} < +0.2$, and an age of $0.7$--3.1~Gyr and the two
populations combined have a mass of $\log{M_*}\sim7.3$--7.8
(Table~\ref{tab:SED}). Thus, the two-population models find younger
ages and larger masses than the single-burst models, and not just
because they allow solutions with higher metallicity than the
single-burst models do.

Since the Lick index absorption-line fits and the single-burst and
two-population SED models yield results in good agreement with one
another we combine them to yield a conservative range of the
parameters of interest that satisfy all of the individual
constraints. For the parameters of primary interest we find that the
full range of acceptable values for the most recent star formation
episode in \dwarf\ are: 0.7--3.4~Gyr age at a metallicity of $-1.7 <
{\rm [Fe/H]} < +0.2$. Of secondary importance, we estimate the stellar
mass of \dwarf\ to be $7.3 < \log{M_*} < 7.8$ and suggest that if more
than one stellar population is present in \dwarf\ then the most recent
one contains $<60$\% of the total galaxy mass.

Using the relative likelihood fractions at each metallicity grid point
(Table~\ref{tab:SED}), the most likely model for \dwarf\ has a stellar
mass of $\log{M_*}\sim7.5$ and $\sim10$\% of this mass is contained in
the galaxy's youngest population, which has an age and metallicity of
$\sim2$~Gyr and ${\rm [Fe/H]}\sim-1$, respectively. We use these
parameters in the next Section to consider whether the assertion of
Paper~1 that \dwarf\ could plausibly produce the 1585~\kms\ absorber
in the 3C~273 sight line still holds in light of our new results.

\section{Updated Absorber Formation Model}
\label{model}

In Paper~1 we modeled the 1585~\kms\ absorber in the 3C~273 sight line
as part of a thin shell of material that was ejected from \dwarf\
approximately 3.5~Gyr ago during a single massive
($\sim10^8~M_{\Sun}$) starburst episode. We concluded that the
supernovae formed in such a starburst are capable of ejecting all of
the galaxy's gas at the time of the starburst \citep[estimated to be
$M_{\rm H\,I} \sim 5\times10^8~M_{\Sun}$;][]{bruzual93,rosenberg02} to
the location where the absorber is observed today. However, the
observations and analysis of Sections~\ref{apo} and \ref{galex} find a
considerably larger recession velocity and younger starburst age for
\dwarf\ than the values assumed in Paper~1. Here we consider whether
simple models of the absorber's origin in a galactic wind emanating
from \dwarf\ are still feasible.

We emphasize that there are many possible origins of the absorbing gas
in the 3C~273 sight line. As discussed in Section~\ref{intro},
NGC~4409 is $\sim1.5$ virial radii away from 3C~273 and has a smaller
line-of-sight galaxy/absorber velocity difference and a considerably
larger luminosity than \dwarf. 3C~273 also lies in the southern
extension of the Virgo cluster and the absorber may originate in
filamentary substructure in the intracluster medium \citep{yoon12}, or
even be affected by hot intragroup gas in this vicinity
\citep{stocke14}. We return to the issue of the galaxy's environment
in Section~\ref{model:environment}.

Our aim in presenting the following analysis is not to convince the
reader that \dwarf\ \textit{must have} produced the absorber in the
3C~273 sight line but to explore whether it plausibly \textit{could
have}. This is an interesting question in its own right because of the
faint luminosity of \dwarf\ \citep[$L\sim0.015\,L^*$;][]{stocke13} and
the large number of galaxies at similarly faint luminosities; the
relationship of quasar absorption line systems to dwarf galaxies is
still very much an open question (see Paper~1) and it is only at the
very lowest redshifts that galaxy redshift surveys are sensitive
enough to detect a representative sample of them.

Paper~1 assumed that the 3C~273 sight line intersects the very edge of
an expanding spherical shell since only one absorption component was
observed (two would be expected if the sight line intersected the
``interior'' of a shell with unity covering fraction) and the measured
line-of-sight velocity difference between the absorber and the galaxy
($\Delta v_{\rm los} = 50\pm50$~\kms) was consistent with the gas
expanding entirely in the plane of the sky at the location of
3C~273. This simplifying assumption no longer holds since our revised
line-of-sight galaxy/absorber velocity difference ($\Delta v_{\rm los}
= 190\pm13$~\kms) is large. Further, we can no longer support the
model of a complete spherical shell since only one absorption
component is observed in \HI\ Lyman series and metal lines
\citep{sembach01,tripp02}; thus, for the spherical shell model of this
absorber to survive, the shell must be patchy. Below we attempt to
constrain the properties of \dwarf\ and its patchy shell of ejecta.

\subsection{Ballistic Wind Model}
\label{model:ballistic}

In Section~\ref{galex:summary} we presented a synthesis of the
absorption-line and SED modelling of \dwarf, determining that
$\sim10$\% of the stars (by mass) were formed $\sim2$~Gyr ago and that
the total stellar mass of old and young stars in the galaxy is
$\log{M_*} \sim 7.5$. Here we use these values to examine a simple
ballistic model of an energy-conserving wind emanating from \dwarf.

\subsubsection{Energy-conserving Wind}
\label{model:ballistic:wind}

\begin{figure}[!t]
\epsscale{1.00} \centering \plotone{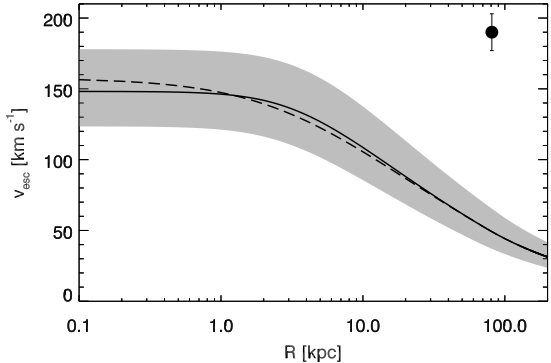}
\caption{Escape speed as a function of radius for a cored dark matter
  profile (solid black line) and an NFW profile (dashed black line)
  with a halo mass of $\log{M_h} = 10.36$. The filled gray region
  shows the variation in escape speed for the cored profile when the
  halo mass varies by $\pm0.25$~dex. The errors on the NFW profile are
  comparable but omitted for clarity. The data point shows the
  observed line-of-sight galaxy/absorber velocity difference ($\Delta
  v_{\rm los}$) and galaxy impact parameter ($\rho$).
\label{fig:vesc}}
\end{figure}

The relations of \citet*{moster13} suggest that \dwarf\ has a halo
mass of $\log{M_h} = 10.36\pm0.25$; if we assume that the halo mass is
equal to the virial mass, then \dwarf\ has a virial radius of $R_{\rm
vir} \approx 75\pm15$~kpc \citep{stocke13}. We have used two different
halo mass distributions to estimate the mass profile of \dwarf: a
cored mass distribution from \citet{salucci07} and the ``cuspy'' NFW
profile \citep*{navarro96}. The escape speed as a function of radius
for these profiles is shown in Figure~\ref{fig:vesc}. While the NFW
profile (dashed black line) has a larger escape speed for very small
radii, as expected, the difference between the two models is never
more than 10~\kms. The two models are identical at $R\gtrsim80$~kpc
because the distributions are truncated to ensure that $M_h = M_{\rm
vir} \equiv M(R_{\rm vir})$ as in \citet{stocke13}.

\begin{figure}[!t]
\epsscale{1.00} \centering \plotone{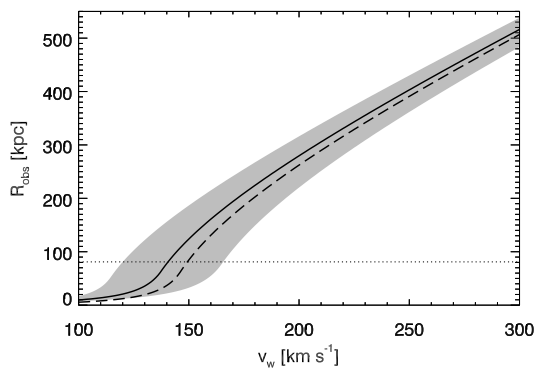}
\caption{Distance the ejecta have traveled in time $t_{\rm sb}=2$~Gyr
  if they had an initial velocity of $v_w$ in an energy-conserving
  wind. The solid and dashed lines are defined as in
  Figure~\ref{fig:vesc}. The horizontal dotted line shows the impact
  parameter, $\rho$, which places a firm lower limit on $R_{\rm obs}$.
\label{fig:vw}}
\end{figure}

The observed radial velocity difference between the absorber and
\dwarf\ is also displayed in Figure~\ref{fig:vesc} at the galaxy
impact parameter ($\rho=81$~kpc at the revised recession velocity of
\dwarf). These values are both projected quantities, and thus lower
limits to the proper three-dimensional values. The \textit{observed}
velocity difference is comparable to the escape speed at $R<1$~kpc,
consistent with \dwarf\ producing an unbound wind as the result of its
most recent starburst.

We utilize conservation of energy to model the wind's velocity as a
function of radius:
\begin{equation}
v^2(R) = v_w^2 - v_{\rm esc}^2(R_0) + v_{\rm esc}^2(R),
\label{eqn:vR}
\end{equation}
where $v_w \equiv v(R_0)$ is the initial wind
velocity. Figure~\ref{fig:vesc} shows that the escape speed is nearly
constant at $R\lesssim1$~kpc so we adopt $v_{\rm esc}(R_0) =
150\pm30$~\kms\ and $R_0 = 0.1$~kpc.

 For the wind model to be consistent with the absorption-line and SED
 models of \dwarf\ it must satisfy
\begin{equation}
t_{\rm sb} = \int_{R_0}^{R_{\rm obs}} \frac{dR}{v(R)},
\label{eqn:tsb}
\end{equation}
where $t_{\rm sb} \sim 2$~Gyr is the age of the starburst and $R_{\rm
obs}$ is the total (i.e., deprojected) galaxy/absorber separation at
the time of observation.

For any value of $v_w$ there is a unique value of $R_{\rm obs}$ that
satisfies Equation~\ref{eqn:tsb}, which we solve for numerically.
This relationship is shown in Figure~\ref{fig:vw}, where the solid and
dashed lines are defined as in Figure~\ref{fig:vesc} and the
horizontal dotted line shows the impact parameter, $\rho$. Since
$R_{\rm obs} \geq \rho$ by definition, we constrain $v_w \gtrsim
170$~\kms\ for the ejecta to reach its observed location.

\subsubsection{Galaxy/Absorber Kinematics}
\label{model:ballistic:kinematics}

We now have enough information to construct a simple model of the
galaxy/absorber geometry. We define $\theta$ to be the angle with
respect to the line-of-sight that describes the absorber trajectory,
where $\theta=0\degr$ if the absorber is moving entirely along the
line-of-sight and $\theta=90\degr$ if the absorber is moving entirely
in the plane of the sky.  The deprojected galaxy/absorber velocity
difference and physical separation between the galaxy and absorber are
then given by
\begin{equation}
\Delta v_{\rm obs} = \frac{\Delta v_{\rm los}}{\cos{\theta}}
\label{eqn:vobs}
\end{equation}
and
\begin{equation}
R_{\rm obs} = \frac{\rho}{\sin{\theta}},
\label{eqn:Robs}
\end{equation}
respectively.

\begin{figure}[!t]
\epsscale{1.00} \centering \plotone{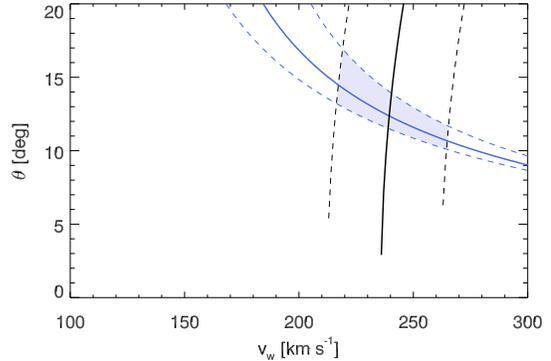}
\caption{Two different constraints on the relationship between $v_w$
  and $\theta$ using a cored mass distribution \citep{salucci07}. The
  solid black line shows the constraint given by
  Equation~\ref{eqn:costheta}, and the solid blue line shows the
  constraint given by Equations~\ref{eqn:tsb} and \ref{eqn:Robs}. The
  dashed lines show the results of varying the halo mass by
  $\pm0.25$~dex for each constraint.  The shaded region is the
  concordance zone where the two constraints agree.
\label{fig:theta}}
\end{figure}

These equations can be used to constrain the relationship of $v_w$ and
$\theta$ in two different ways.  First, we recognize that $\Delta
v_{\rm obs} \equiv v(R_{\rm obs})$ and use Equations~\ref{eqn:vR} and
\ref{eqn:vobs} to find:
\begin{equation}
\cos{\theta} = \Delta v_{\rm los} \left[ v_w^2 - v_{\rm esc}^2(R_0) +
v_{\rm esc}^2(\rho) \right]^{-1/2}.
\label{eqn:costheta}
\end{equation}
Here we have approximated $R_{\rm obs}$ with $\rho$ in the final term
under the square root. Since the value of the escape velocity is
relatively small at $R>\rho$ (i.e., $v_{\rm esc}(\rho) =
50\pm15$~\kms; Figure~\ref{fig:vesc}) this substitution has little
effect on the final result and leaves $v_w$ as the only unknown on the
right-hand side of Equation~\ref{eqn:costheta}. This relationship is
shown as the solid black line in Figure~\ref{fig:theta}.

Similarly solving for $\sin{\theta}$ as a function of $v_w$ by
combining Equations~\ref{eqn:tsb} and \ref{eqn:Robs} yields the solid
blue line in Figure~\ref{fig:theta}.  Unfortunately, we cannot derive
an analytic expression for this relationship since we solve
Equation~\ref{eqn:tsb} numerically for the value of $R_{\rm obs}$ at
which the integral equals a given $t_{\rm sb}$.    The dashed lines in
Figure~\ref{fig:theta} show the propagated uncertainties associated
with varying the halo mass by $\pm0.25$~dex, and  the shaded region
indicates where the two constraints agree: $\theta =
12^{+5}_{-2}$~degrees and $v_w = 240\pm25$~\kms. This initial wind
velocity is large but still reasonable for a supernova-driven wind
\citep{maclow99,martin99} in either a large or small galaxy; i.e., the
supernovae blast wave velocity is independent of galaxy mass and
luminosity.

The concordance range of $\theta$ yields $R_{\rm obs} =
390^{\phn+70}_{-110}$~kpc and $\Delta v_{\rm obs} = 195\pm15$~\kms,
which are $\sim5$ times $R_{\rm vir}$ and $\sim9$ times $v_{\rm
esc}(R_{\rm  obs})$, respectively. These results suggest that if
\dwarf\ formed a galaxy wind as a result of its most recent episode of
star formation $\sim2$~Gyr ago, then any ejecta associated with that
wind have escaped the galaxy easily and traveled to very large
distances if unimpeded.

\subsection{The Gaseous Shell Surrounding \dwarf}
\label{model:shell}

Now that the galaxy/absorber kinematics have been examined in the
context of an energy-conserving galaxy wind we turn our attention to
estimating the amount of material that could have been ejected from
dwarf given its initial wind velocity
(Section~\ref{model:shell:energetics}) and the amount of material that
can be inferred to surround \dwarf\ if the 1585~\kms\ absorber in the
3C~273 sight line is part of a shell of ejecta
(Section~\ref{model:shell:ionization}).

\subsubsection{Mass Estimate from Wind Energetics}
\label{model:shell:energetics}

Our synthesis model (Section~\ref{galex:summary}) suggests that
\dwarf\ formed $\sim3\times10^6~M_{\Sun}$ of stars (10\% of
$\log{M_*}=7.5$) in a starburst $\sim2$~Gyr ago. Using a Salpeter IMF
from 0.1--100~$M_{\Sun}$ this corresponds to $\sim2\times10^4$ stars
with $M > 8~M_{\Sun}$ that will create supernovae. Assuming
$10^{51}$~erg of kinetic energy per supernova was converted to bulk
motion with an efficiency of 3--30\% \citep{koo92a,koo92b,cioffi91}
the starburst provided $\log{(E_k/{\rm erg})} \approx 53.9$--54.9 of
kinetic energy to power a galactic wind.  Fewer young stars are
created in this model compared to the value assumed in Paper~1
($\geq10^8~M_{\Sun}$), so our estimate of the available kinetic energy
is $\sim1.4$~dex less than previously.

Combining the kinetic energy that the starburst injected with the
initial wind velocity from our kinematic model suggests that
\begin{equation}
M_{\rm  shell} = \frac{2E_k}{v_w^2}
\label{eqn:Mshell:Ek}
\end{equation}
of gas could have been entrained in a wind and ejected from
\dwarf. Using the value of $v_w=240\pm25$~\kms\ from
Section~\ref{model:ballistic:kinematics} suggests that $M_{\rm shell}
\sim (7\pm4)\times10^6~M_{\Sun}$, which is more than the total mass of
stars created in the burst.

\subsubsection{Mass Estimate from Absorber Properties}
\label{model:shell:ionization}

In this subsection we examine the ionization conditions of the
1585~\kms\ absorber near the 3C~273 sight line with the goal of
estimating the density of the absorbing material. When combined with
the observed column density this density gives an estimate of the
absorber's line-of-sight thickness, which can be used to constrain the
thickness of any shell of material that may have been ejected from
\dwarf, and subsequently to estimate the amount of mass contained in
that shell.

{\sl Far Ultraviolet Spectroscopic Explorer} ({\sl FUSE}) observations
of 3C~273 revealed seven high-order Lyman series lines
(Ly$\beta$--Ly$\theta$) that constrain the \HI\ column density and
Doppler parameters of the 1585~\kms\ absorber to be $\log{N_{\rm H\,I}
= 15.85^{+0.10}_{-0.08}}$ and $b_{\rm H\,I} = 16\pm1$~\kms,
respectively \citep{sembach01}. \citet{tripp02} combined these {\sl
FUSE} observations with a high signal-to-noise echelle spectrum from
the Space Telescope Imaging Spectrograph (STIS) aboard \hst\ to
construct detailed photoionization models of the absorber. These
CLOUDY models \citep{ferland98} indicate that the absorber has a
metallicity of $[{\rm C/H}] = -1.2^{+0.3}_{-0.2}$, a silicon
overabundance of $[{\rm Si/C}] = +0.2\pm0.1$, a density of
$\log{n_{\rm H}} = -2.8\pm0.3$, and a line-of-sight thickness of $D =
N_{\rm H}/n_{\rm H} = 70^{+280}_{-60}$~pc \citep{tripp02}.

\begin{deluxetable}{lcc}

\tablecolumns{3}
\tablewidth{0pt}

\tablecaption{\hst/STIS and COS Column Densities
\label{tab:absorber}}

\tablehead{ \colhead{Species} & \colhead{$\log{N_{\rm STIS}}$} & \colhead{$\log{N_{\rm COS}}$} }

\startdata
C\,{\sc ii}   & $12.74^{+0.10}_{-0.12}$ & $12.85\pm0.13$ \\
C\,{\sc iv}   & $<12.5$                 & $<12.13$       \\
Si\,{\sc ii}  & $11.76^{+0.11}_{-0.15}$ & $11.87\pm0.11$ \\
Si\,{\sc iii} & $12.33\pm0.08$          & $12.22\pm0.09$ \\
Si\,{\sc iv}  & $<12.0$                 & $<11.74$       
\enddata

\tablecomments{All column densities are in units of ${\rm cm}^{-2}$. STIS values are from Table~1 of \citet{tripp02}. All upper limits are quoted to $3\sigma$ confidence.}

\end{deluxetable}

3C~273 was also observed with the Cosmic Origins Spectrograph aboard
\hst\ on 2012~Apr~22 as part of GTO program 12038 (PI: J.~Green). The
data were reduced and absorption line properties measured as detailed
in \citet{keeney13} and \citet{stocke13}. Table~\ref{tab:absorber}
shows the ionic column densities used in the \citet{tripp02} analysis
compared to those found in the COS spectrum.

\begin{figure}[!t]
\epsscale{1.00} \centering \plotone{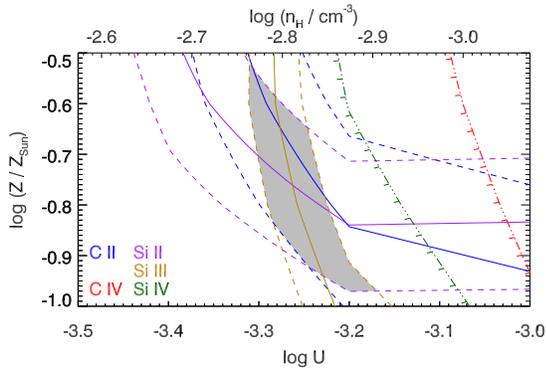}
\caption{Photoionization model generated by the COS column densities
  in Table~\ref{tab:absorber}. The solid lines show the observed
  column densities and the dashed lines show the quadrature sum of the
  ionic column density error and the \HI\ column density error. The
  dot-dashed lines show upper limits with the tick marks indicating
  the allowed region of parameter space. The shaded region shows where
  the observed \ion{Si}{2}, \ion{Si}{3}, and \ion{C}{2} column
  densities can be simultaneously reproduced.
\label{fig:cloudy}}
\end{figure}

Assuming the \HI\ column density of \citet{sembach01}, we have
generated a photoionization model of the 1585~\kms\ absorber using the
COS values from Table~\ref{tab:absorber} and the same methodology as
in \citet{stocke13}. The absorber was modeled as a plane-parallel slab
illuminated by the extragalactic ionizing background of
\citet{haardt12}. We assumed that no ionizing flux from hot stars in
\dwarf\ would reach the absorber location (nor is any expected; see
Figure~\ref{fig:SED}). This new model is shown in
Figure~\ref{fig:cloudy} and finds $\log(Z/Z_{\Sun}) =
-0.8^{+0.3}_{-0.2}$, $\log{n_{\rm H}} = -2.8\pm0.1$, and $D =
130^{+45}_{-30}$~pc. Unlike the \citet{tripp02} analysis, we have
assumed relative solar abundances and only varied the metallicity,
$Z$, and ionization parameter, $U$. While this photoionization model
and the model of \citet{tripp02} vary in their details, they infer
very similar properties for the absorbing gas. We note, however that
the metallicity in our new model is higher than the value found by
\citet{tripp02}.

Despite the many similarities in absorber properties between our
models and those of \citet{tripp02}, there are systematic
uncertainties in our results. We have not explored the possibility
that the absorber is collisionally ionized, but we think that this is
unlikely due to its relatively high \HI\ column density and
non-detection in any ion more highly ionized than \ion{Si}{3}. We have
modelled all of the observed ions as arising in the same phase of gas,
which we feel is relatively safe given the low ionization of all the
detected species (\HI, \ion{C}{2}, \ion{Si}{2}, \ion{Si}{3}) and their
velocity alignment in the STIS data \citep[which has a more stable
wavelength solution than COS;][]{tripp02}. Other limitations of our
simple models are that we have not considered non-equilibrium effects,
nor have we explored different UV background shapes to see how they
affect our results.

Both our model and that of \citet{tripp02} infer a density of $n_{\rm
H} \sim 1.6\times10^{-3}~{\rm cm}^{-3}$ and a line-of-sight  thickness
of $D\sim100$~pc for the 1585~\kms\ absorber toward 3C~273. If we
assume that ejecta from \dwarf\ formed a fully-filled spherical shell
of uniform density and thickness $dR \approx D\cos{\theta}$ and radius
$R_{\rm obs}$ then the shell has a mass of
\begin{align}
\label{eqn:Mshell:Robs}
M_{\rm shell} &= m_{\rm H} n_{\rm H} V_{\rm shell} \\ &= \frac{4\pi
              m_{\rm H} n_{\rm H}}{3} \left[R_{\rm obs}^3 - (R_{\rm
              obs}-dR)^3\right]. \nonumber
\end{align}
Using the values of $R_{\rm obs}$ and $\theta$ from
Section~\ref{model:ballistic} yields $M_{\rm shell}
\sim10^{10}~M_{\Sun}$, much larger than the supernovae associated with
the last burst of star formation in \dwarf\ were capable of producing
(Section~\ref{model:shell:energetics}).

The varying estimates of the shell mass from
Equations~\ref{eqn:Mshell:Ek} and \ref{eqn:Mshell:Robs} imply that the
shell of ejecta is partially covered, with $f_c\ll1$\% of its surface
containing material similar to that which we observe toward the 3C~273
sight line if all of the observed material originated in \dwarf. This
fraction is so small that it is unlikely that a ballistic wind from
\dwarf\ created the 1585~\kms\ absorber.

\subsection{The Effect of Galaxy Environment}
\label{model:environment}

Section~\ref{model:ballistic} specifies the simplest possible wind
model, where the galaxy is treated in isolation and the ejecta do not
interact with any other material after the time of ejection. It
therefore presents a limiting case of the farthest that the galaxy
could have ejected material in the time since the last episode of star
formation.

We know that \dwarf\ resides in a rich galaxy environment, however,
with ample gaseous material with which wind ejecta can interact
\citep{yoon12,stocke14}. Such interactions would slow down the ejecta
with respect to the ambient medium. The galaxies in this part of the
Virgo cluster have a velocity centroid of $\sim1600$~\kms\
\citep{yoon12,stocke14}, similar to the velocity of the 3C~273
absorber. Given that the velocity of \dwarf\ is 1775~\kms, if we
assume that the ambient gaseous medium has a similar velocity to the
surrounding galaxies then the interaction of any ejecta with this
medium would cause the measured value of $\Delta v_{\rm los}$ to be
larger than the value associated with the initial wind ejection,
$\Delta v_{\rm los,wind}$. Similarly, if the ejecta do not travel
ballistically for the entire time between ejection and observation
then the relevant time to use in Equation~\ref{eqn:tsb} is not $t_{\rm
sb}$ but rather $t_{\rm ball}$, the time that the ejecta travelled
ballistically.

\begin{figure}[!t]
\epsscale{1.00} \centering \plotone{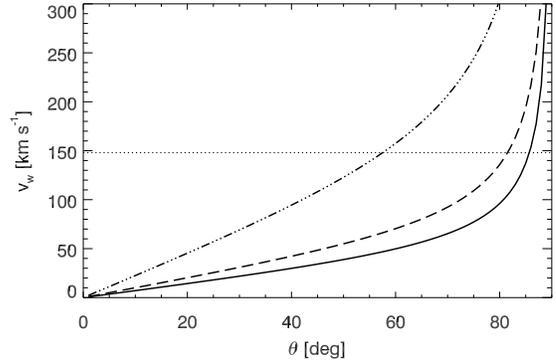}
\caption{The relationship between $v_w$ and $\theta$
  (Equation~\ref{eqn:fc}) for shell covering fractions of $f_c=1$
  (solid line), $f_c=0.5$ (dashed line), and $f_c=0.1$ (dash-dotted
  line). The dotted horizontal line shows the minimum value of $v_w$
  that allows the shell to escape \dwarf.
\label{fig:fc}}
\end{figure}

Combining Equations~\ref{eqn:Mshell:Ek} and \ref{eqn:Mshell:Robs} and
accounting for partial covering yields
\begin{equation}
v_w^2 = \frac{3E_k}{2\pi m_{\rm H} n_{\rm H} f_c} \left[
\left(\frac{\rho}{\sin{\theta}}\right)^3 -
\left(\frac{\rho}{\sin{\theta}} - D\cos{\theta}\right)^3 \right]^{-1}.
\label{eqn:fc}
\end{equation}
This relationship is shown in Figure~\ref{fig:fc} for several values
of $f_c$. For the shell to escape \dwarf\ we require $v_w > v_{\rm
esc}(R_0)$ (horizontal dotted line; see Figure~\ref{fig:vesc}), which
sets limits of $\theta>84\degr$ ($f_c=1$), $\theta>78\degr$
($f_c=0.5$), and $\theta>50\degr$ ($f_c=0.1$). Thus, the absorber
trajectory must be close to the plane of the sky for any plausible
value of $f_c$.
 
The most plausible models are those where $f_c\gtrsim0.5$, for which
$\theta\gtrsim80\degr$ and $R_{\rm obs}\approx\rho$ (i.e., the values
assumed in Paper~1); substituting $\rho$ for $R_{\rm obs}$ in
Equation~\ref{eqn:tsb} yields $t_{\rm ball} \lesssim
1.5$~Gyr. Regardless of the value of $f_c$,
Equation~\ref{eqn:Mshell:Ek} indicates that the shell mass can be no
more than $2E_k/v_{\rm esc}^2(R_0) \sim
(3.5\pm1.2)\times10^7~M_{\Sun}$, which is about 10 times more than the
mass of stars formed in the most recent burst in \dwarf.

\subsection{Summary of Wind Models}
\label{model:summary}

In this Section we have explored whether \dwarf\ is capable of
producing the absorption seen at 1585~\kms\ in the 3C~273 sight
line. In Section~\ref{model:ballistic} we developed an
energy-conserving wind model and attempted to model the
galaxy/absorber kinematics in the context of this model. We assumed
that \dwarf\ last formed stars $t_{\rm sb}\sim2$~Gyr ago and the
resultant supernovae explosions created a wind that is responsible for
the observed line-of-sight galaxy/absorber velocity difference of
$\Delta v_{\rm los} = 190\pm13$~\kms. We further assumed that this
wind has travelled ballistically for the entire time $t_{\rm sb}$
since formation and that it is responsible for the current lack of gas
in \dwarf\ \citep{vangorkom93}; i.e., we assumed that \dwarf\ is
solely responsible for the formation of the absorber.

Under these assumptions the wind emanating from \dwarf\ must intersect
the 3C~273 sight line at an angle of $\theta \approx 10\degr$ with
respect to the line of sight and had an initial velocity of $v_w
\approx 240$~\kms. Deprojecting the observed galaxy/absorber
separation on the sky ($\rho=81$~kpc) and velocity difference ($\Delta
v_{\rm los}$) using Equations~\ref{eqn:vobs} and \ref{eqn:Robs} yields
$R_{\rm obs} \approx 5\rho$ and $\Delta v_{\rm obs} \approx \Delta
v_{\rm los}$. Thus, this ballistic model is the limiting case of the
farthest that \dwarf\ could have expelled material in a time $t_{\rm
sb}$; as such, it is a ``maximal expansion'' model where the
three-dimensional velocity is projected almost entirely onto the line
of sight.

In Section~\ref{model:shell} we considered the implications of this
ballistic wind model for the gaseous shell of material surrounding
\dwarf, of which the absorber is a part under our assumptions. We find
that the supernovae explosions associated with the last burst of star
formation in \dwarf\ would provide enough kinetic energy to expel a
shell of gas with a mass comparable to the mass of stars formed in the
burst. However, a thin spherical shell with uniform density comparable
to that which we derive from the 3C~273 absorber ($n_H \sim
1.6\times10^{-3}~{\rm cm}^{-3}$) and radius $\sim5\rho$ would have a
much larger mass that is comparable to the galaxy's entire halo
mass. Thus, we reject the simple ballistic wind model for the
formation of the absorber due to its untenably low covering fraction.

In Section~\ref{model:environment} we surmised that one consequence of
the rich galaxy environment near \dwarf\ is that the observed $\Delta
v_{\rm los}$ may not be indicative of the initial wind velocity due to
interactions of the wind ejecta with the ambient gaseous medium. We
develop a relation between the initial wind velocity and trajectory
for a shell of a given covering fraction, $f_c$
(Equation~\ref{eqn:fc}), and find that for $f_c\gtrsim0.5$ the wind
velocity is larger than the galaxy's escape velocity only for
$\theta\gtrsim80\degr$, which implies $R_{\rm obs}\approx\rho$. This
high-covering-fraction model thus represents a ``minimal expansion''
model in which the galaxy/absorber separation is projected almost
entirely onto the plane of the sky.

The only robust conclusion that we can draw from this analysis is that
\dwarf\ cannot be solely responsible for the formation of the
1585~\kms\ absorber in the 3C~273 sight line. The ballistic wind model
explicitly tried to account for the absorber by an escaping wind from
\dwarf\ alone and failed because the supernovae associated with the
galaxy's most recent star formation episode had insufficient kinetic
energy to fill a large fraction of the surface of an expanding shell
with material similar to that which we observe toward 3C~273. The
high-covering-fraction model invokes \textit{a priori} the presence of
ambient gas with which the wind ejecta interact to decouple the
observed $\Delta v_{\rm los}$ and the initial wind velocity. Since
these two models represent different extremes of wind trajectory and
physical extent and both require wind ejecta to interact with gas
outside of \dwarf\ to be tenable, we expect any plausible model at
intermediate trajectories will require this external gas as well.

\section{Summary and Conclusions}
\label{conclusion}

In Paper~1 we presented an optical spectrum and absorption-line
analysis of the dwarf post-starburst galaxy \dwarf\ and argued that a
supernova-driven wind from this galaxy could explain the formation of
the 1585~\kms\ absorber in the nearby 3C~273 sight line. In this Paper
we present new high-resolution optical spectroscopy and \galex\ images
of \dwarf, as well as new \hst/COS spectroscopy of 3C~273, to further
explore the association of this galaxy/absorber pair.

We presented a new high-resolution optical spectrum of \dwarf\ in
Section~\ref{apo}. This spectrum confirms the galaxy redshift found by
SDSS, which is $\sim150$~\kms\ higher than the value reported in
Paper~1. It also confirms the lack of \Ha\ emission in \dwarf, and
both the \Ha\ and FUV data confirm that its current SFR is
$<10^{-3}~M_{\Sun}\,{\rm yr}^{-1}$. The galaxy's metallicity and the
age of its most recent star formation episode were determined from the
SDSS spectrum and found to be consistent with the values obtained for
the single-burst SED models of Section~\ref{galex} and the Lick index
analysis of Paper~1. Thus, the conclusion that this dwarf will not
continue to form new stars but rather fade to a very low luminosity
($M_{\rm B} \geq -14$) is confirmed.

In Section~\ref{galex} we examined the impact of both single-burst and
two-population models for the star formation history of \dwarf\ on the
galaxy's UV-optical SED. The \galex\ UV magnitudes were essential for
distinguishing between models of different ages.  We find that both
single-burst and two-population star formation histories provide
acceptable fits to the galaxy's SED, but the best-fit two-population
models have significantly smaller reduced $\chi^2$ values than the
best-fit single-burst models (see Figure~\ref{fig:SED}). The
absorption-line fits and SED analysis yield comparable estimates for
the age and metallicity of the youngest stellar population in
\dwarf. We combine these results to find a \textit{full range} of
acceptable values for the age, metallicity, and mass fraction of this
population of 0.7--3.4~Gyr, $-1.7 < {\rm [Fe/H]} < +0.2$, and
$f_m<0.6$, respectively, and estimate the total stellar mass of
\dwarf\ to be $7.3 < \log{M_*} < 7.8$.

In Section~\ref{model} we attempted to synthesize these results into a
simple model of the galaxy/absorber system to confirm whether the
conclusion of Paper~1 that a wind emanating from \dwarf\ could
plausibly form the 3C~273 absorber is still valid given our new
data. Given the broad range of acceptable star formation histories for
\dwarf, we choose fiducial values of $f_m\sim0.1$, $\log{M_*}\sim7.5$,
$t_{\rm sb}\sim2$~Gyr, and ${\rm [Fe/H]}\sim-1$ for this analysis.  We
explore both minimal and maximal expansion models where the
three-dimensional galaxy/absorber separation is projected to lie
almost entirely on the plane of the sky or along the line of sight,
respectively, and find that any wind ejecta from \dwarf\ must interact
with gas outside of the galaxy to explain the presence of the 3C~273
absorber in both cases. Therefore, we conclude that \dwarf\ cannot be
solely responsible for the formation of the absorber.

The properties of \dwarf\ that led us to speculate in Paper~1 that it
could plausibly form a starburst wind are its post-starburst optical
spectrum and its present-day lack of gas \citep{vangorkom93}. While
these properties remain unchanged, we no longer find compelling
evidence that the 1585~\kms\ absorber in the 3C~273 sight line was
formed solely by a wind escaping from \dwarf. We have revised the
recession velocity of \dwarf\ upward from the value presented in
Paper~1, however, and argue in Section~\ref{model:environment} that
the galaxies in the southern extension of the Virgo cluster have a
centroid of $\sim1600$~\kms, comparable to the absorber velocity
\citep{yoon12,stocke14}. If we assume that any intracluster gas in the
vicinity has a similar velocity then \dwarf\ is moving through this
medium with a relative velocity of $\sim175$~\kms. This velocity is
high enough that we may not require supernovae associated with the
last burst of star formation in \dwarf\ to explain its present-day
lack of gas (i.e., ram pressure stripping may be a viable alternative).

Another consequence of the potentially high peculiar velocity of
\dwarf\ is that it may have moved an appreciable distance \textit{on
the sky} since its last burst of star formation $\sim2$~Gyr ago. If
\dwarf\ has a velocity of $\sim100$~\kms\ in the plane of the sky then
it has travelled $\sim200$~kpc on the sky ($\sim28\arcmin$,
corresponding to a proper motion of $\sim10^{-3}~{\rm mas\,yr^{-1}}$)
in the past 2~Gyr. Thus, the current proximity of \dwarf\ to the
3C~273 sight line may be coincidental as it could be caught in the act
of passing by. This is a fundamental ambiguity in trying to associate
QSO absorption line systems with individual galaxies since there can
be a large physical distance between the galaxy and the QSO sight line
and the galaxy has a proper motion that is too small to be measured;
thus, there is no way to know the relative positions of the galaxy and
the QSO sight line several Gyr in the past when the absorbing material
was presumably expelled from the galaxy.

It is instructive to look at the galaxy/absorber pair studied here in
the larger context of what is being learned about the circumgalactic
medium (CGM) of galaxies from COS observations. Targeted and
serendipitous galaxies located within $\sim0.5\,R_{\rm vir}$ of the
QSO sight line typically show low- or intermediate-ionization
absorption from \HI\ and metal-lines that are well-fit by
photoionization models similar to those presented in
Section~\ref{model:shell:ionization}
\citep{prochaska11,tumlinson11,werk13,werk14,stocke13}. These close-in
CGM systems often have strong collisionally-ionized \ion{O}{6}
absorption as well \citep{tumlinson11,stocke13,stocke14}.

The situation is less clear for galaxies located $\gtrsim1\,R_{\rm
vir}$ from the QSO sight line (i.e., those with similar separations to
\dwarf\ and 3C~273), however. At these distances the incidence of
\lya\ absorption is similar to that for the closer galaxies, but the
incidence of metals is not. Metals are not detected for $\sim50$\% of
galaxies with $\rho \sim 1\,R_{\rm vir}$, and when they are detected
it is typically only in higher-ionization species such as \ion{O}{6} 
and \ion{C}{4} \citep[e.g., see Table~3 and Figures~8 and 11
of][]{stocke13}. Therefore, the connection between individual galaxies
located $\sim1\,R_{\rm vir}$ from a QSO sight line and the detected
QSO absorption-line system is ambiguous. While the relatively low
ionization of the 1585~\kms\ absorber in the 3C~273 sight line and the
unusual properties of \dwarf\ compared to other pairs at these
separations made it seem more likely that a definitive association
between the two could be made, we have not been able to make a strong
case for a causal relationship.

The Q1230+0115 sight line is located near 3C~273 on the sky
($\sim1\degr$ separation) and has an absorption-line system at
comparable velocity \citep[1700~\kms;][]{rosenberg03} and \HI\ column
density \citep[$\log{N_{\rm H\,I}} = 15.27\pm0.22$;][]{tilton12} to
the 3C~273 absorber studied here \citep[1585~\kms\ and $\log{N_{\rm
H\,I}} = 15.85^{+0.10}_{-0.08}$, respectively;][]{sembach01}. The
Q1230+0115 absorber is also detected in several metal species,
including \ion{C}{2}, \ion{C}{4}, \ion{Si}{2}, and \ion{Si}{4}
\citep{rosenberg03}. The region around Q1230+0115 has been surveyed
for galaxies to comparable depth to the region around 3C~273. The
galaxy that is the fewest number of virial radii from the 1700~\kms\
absorber is NGC~4536, which has $\rho = 550~{\rm kpc} \approx
2.1\,R_{\rm vir}$ and $\Delta v_{\rm los} \approx 100$~\kms. There are
several smaller galaxies within 400~\kms\ of the absorber redshift
that are located closer to the sight line, but they all have
$\rho>240$~kpc. Thus, it is unlikely that a single galaxy is
responsible for producing the metals observed in this absorber either.

We have been unable to find galaxies located closer than
$\sim1\,R_{\rm vir}$ to very low redshift metal-line absorbers in the
3C~273 and Q1230+0115 sight lines despite surveying to very faint
levels \citep[$L < 0.001\,L^*$;][]{stocke13} in both cases. In this
Paper we concluded that the dwarf post-starburst galaxy \dwarf\ could
not create the 3C~273 absorber on its own and the closest galaxy to
the Q1230+0115 absorber (in terms of virial radii) is twice as far
from the sight line as \dwarf. Thus, these absorbers appear to have
truly intergalactic (i.e., composite), rather than circumgalactic
(i.e., single-galaxy), origins. These results suggest that the only
truly circumgalactic absorbers may be those located within
$\sim0.5\,R_{\rm vir}$ of galaxies \citep[e.g.,][]{stocke13}.

\acknowledgements We wish to thank the referee for careful and
constructive criticisms that have improved the quality of this work.
This effort was supported by NASA grants NNX08AU62G (\galex\ GI
Program) and NNX08AC14G (\hst/COS) to the University of Colorado at
Boulder. BAK and JTS gratefully acknowledge additional support from
NSF grant AST-1109117. PJ acknowledges financial support from the
Center for Astrophysics \& Space Astronomy at the University of
Colorado Boulder.

{\it Facilities:} \facility{GALEX}, \facility{APO (DIS)},
\facility{HST (COS)}

\end{document}